\documentclass[prd, nofootinbib, notitlepage, twocolumn]{revtex4-1}

\usepackage[caption=false]{subfig}
\usepackage{graphicx}
\usepackage{amsmath}

\usepackage{xcolor}
\usepackage{ulem}
\definecolor{naviBlue}{RGB}{0,0,128}
\usepackage[colorlinks  = true,
            linkcolor   = naviBlue,
            urlcolor    = naviBlue,
            citecolor   = naviBlue,
            anchorcolor = naviBlue   ]{hyperref}

\newcommand{\gsim}{\lower.7ex\hbox{$\;\stackrel{\textstyle>}{\sim}\;$}}
\newcommand{\lsim}{\lower.7ex\hbox{$\;\stackrel{\textstyle<}{\sim}\;$}}            

\newcommand{\appref} [1]{\hyperref[app::#1]{Appendix\;\ref*{app::#1}}}
\newcommand{\secref} [1]{\hyperref[sec::#1]{Sec.\;\ref*{sec::#1}}}
\newcommand{\figref} [1]{\hyperref[fig::#1]{Fig.\;\ref*{fig::#1}}}
\newcommand{\tabref} [1]{\hyperref[tab::#1]{Table\;\ref*{tab::#1}}}
\newcommand{\eqnref} [1]{\hyperref[eqn::#1]{Eq.\;(\ref*{eqn::#1})}}

\newcommand{\DM}{\mathrm{DM}}

\newcommand{\pb}{{\bar{p}}}
\newcommand{\nb}{{\bar{n}}}
\newcommand{\db}{{\bar{D}}}
\newcommand{\Heb}{{\overline{\mathrm{He}}}}

\newcommand{\sS}{\sqrt{s}}

\newcommand{\pT}{p_{\mathrm{T}}}

\newcommand{\vk}{\vec{k}}
\newcommand{\pcoal}{p_\mathrm{\coal}}
\newcommand{\coal}{\mathcal{C}}

\newcommand{\sigmaTot}{\sigma_{\mathrm{tot}}}
\newcommand{\sigmaEl }{\sigma_{\mathrm{el }}}

\newcommand{\sv }{\langle \sigma v \rangle}
\newcommand{\mDM}{m_\mathrm{DM}}

\begin{document}

\title{Prospects to verify a possible dark matter hint in cosmic antiprotons with antideuterons and antihelium}

\author{Michael Korsmeier}
\email{michael.korsmeier@to.infn.it}
\affiliation{Dipartimento di Fisica, Universit\`a di Torino, via P. Giuria 1, 10125 Torino, Italy}
\affiliation{Istituto Nazionale di Fisica Nucleare, Sezione di Torino, Via P. Giuria 1, 10125 Torino, Italy}
\affiliation{Institute for Theoretical Particle Physics and Cosmology, RWTH Aachen University, 52056 Aachen, Germany}
\author{Fiorenza Donato}
\email{fiorenza.donato@to.infn.it}
\affiliation{Dipartimento di Fisica, Universit\`a di Torino, Via P. Giuria 1, 10125 Torino, Italy}
\affiliation{Istituto Nazionale di Fisica Nucleare, Sezione di Torino, Via P. Giuria 1, 10125 Torino, Italy}
\author{Nicolao Fornengo}
\email{nicolao.fornengo@to.infn.it}
\affiliation{Dipartimento di Fisica, Universit\`a di Torino, Via P. Giuria 1, 10125 Torino, Italy}
\affiliation{Istituto Nazionale di Fisica Nucleare, Sezione di Torino, Via P. Giuria 1, 10125 Torino, Italy}

\begin{abstract}
\noindent
Cosmic rays are an important tool to study dark matter annihilation in our Galaxy. Recently, a possible hint for 
dark matter annihilation was found in the antiproton spectrum measured by AMS-02,
even though the result might be affected by theoretical uncertainties.
A complementary way to test its dark matter interpretation would be the observation of low-energy antinuclei in cosmic rays.
We determine the chances to observe antideuterons with GAPS and AMS-02, and the implications for the ongoing AMS-02 antihelium searches. We find that the corresponding antideuteron signal are within the GAPS and AMS-02 detection potential.
If, more conservatively, the putative signal was considered as an upper limit on DM annihilation, our results would indicate the highest possible fluxes for antideuterons and antihelium compatible with current antiproton data.

\end{abstract}

\maketitle

\section*{\label{sec::introduction}Introduction}

Astroparticle physics of Galactic cosmic rays (CR) has entered a new level of precision with the measurements of the space borne AMS-02 experiment,
which determined 
proton and helium fluxes    \cite{AMS-02_Aguilar:2015_ProtonFlux, AMS-02_Aguilar:2015_HeliumFlux} at the percent level 
and the antiproton flux     \cite{AMS-02_Aguilar:2016_AntiprotonFlux} at 5\%. Furthermore, the 
B/C ratio                   \cite{Aguilar:2016vqr} and 
lepton fluxes               \cite{2014PhRvL.113l1102A, 2014PhRvL.113v1102A, 2013PhRvL.110n1102A} are now available with unprecedented precision. 
On the other hand, uncertainties in the theoretical models of CR production at the sources and in their propagation are still considerably large. Recent analyses aim to reduce these uncertainties by adapting the models to the new precise AMS-02 data, 
many using Monte Carlo techniques to  properly cover the large space of propagation parameters
\cite{2016PhRvD..94l3019K, Kappl:2015bqa, Evoli:2015vaa, Giesen:2015ufa, Genolini:2015cta, Porter:2017vaa}. 
In this context, a possible hint for dark matter (DM) was found in the AMS-02 antiproton flux  independently by 
 two analyses \cite{Cuoco:2016eej, Cui:2016ppb}. Both of them rely on the numerical tool \textsc{Galprop} \cite{Strong:1998fr} for 
 Galactic propagation 
and find a preference for annihilation of DM particles with a mass between 30 and 100~GeV and with a thermally-averaged cross section close to the WIMP natural scale for a cold thermal relic whose  value is $3\cdot 10^{-26}$~$\mathrm{cm^3/s}$.
These analyses suffer from large systematic uncertainties, and a firm confirmation (or disproval) of this potential signal will require a better understanding of, especially, solar modulation, correlation of uncertainties in the AMS-02 data, Galactic propagation, and the antiproton production cross section. Nevertheless, it appears both compelling and timely to investigate this hint in other detection channels which are going to become experimentally available soon: cosmic antideuterons \cite{Donato:1999gy} and antihelium \cite{Cirelli:2014qia,Carlson:2014ssa}.
In fact, the most direct option to cross-check an antiproton signal is to investigate the associated production of heavier antinuclei, if experimentally accessible: physical processes that lead to the production of antinucleons also lead to the production of heavier antinuclei, even though at a much lower rate. An advantage of the 
DM signal over the secondary background, produced by interaction of CR nuclei on the interstellar medium (ISM), is that the 
latter secondary antimatter production is kinematically strongly suppressed at low energies. 
Therefore, a much more favorable  signal-to-background ratio is expected. 
Antideuterons are the primary goal of the GAPS experiment \cite{Aramaki:2015laa,Aramaki:2015pii}, which is approved by NASA and 
will be launched on balloon within the next few years. Both antideuterons and antihelium are also among the channels of investigations of AMS-02. 

In this paper we investigate what are the implications for the search of antideuterons and antihelium arising from the chance that the fluctuation in the AMS-02 antiproton spectrum is due to DM annihilation. Baseline for our investigations is the analysis by Couco, Kr\"amer, and Korsmeier \cite{Cuoco:2016eej} (in the following CuKrKo), from which we take the particle DM properties compatible to the potential DM hint.
The results will be then confronted with the expected sensitivities of GAPS (for antideuterons) and AMS-02 (for both antideuterons and antihelium).

The paper is structured as follows. In \secref{methods} we describe the coalescence models we adopt to calculate the
astrophysical and DM source terms, and we define the adopted propagation model. Then in \secref{results} we state our results
before concluding in \secref{conclusion}.

\section{\label{sec::methods}Methods}

\subsection{\label{sec::XS}Cross section and coalescence}

The production of cosmic antinuclei by scatterings of CRs off the ISM was first discussed  
in the context of an analytic coalescence model in~\cite{Chardonnet:1997dv}. 
This approach  was then adopted for the production of both secondary and DM originated antinuclei by several groups, see e.g.
\cite{Donato:1999gy, Duperray:2002pj, Duperray:2005si, Ibarra:2009tn, Brauninger:2009pe, Cirelli:2014qia,Carlson:2014ssa}\footnote{Note that the approach in  \cite{Duperray:2002pj} is slightly different. The antideuteron production cross sections are derived in a diagrammatic scheme, which is in rough agreement with the analytic coalescence model \cite{Duperray:2005si}.}.  
In this model, the criterion for coalescence of the individual nucleons is that their relative momentum is small enough to allow the formation of the bound state: this threshold momentum, called the coalescence momentum $\pcoal$, is not known from first principles and has to be determined by adapting the coalescence model to antinuclei production data at accelerators, when available. 

A simple assumption in the early coalescence models is the one of a spherically symmetric antineucleon production: antinucleons are first produced by the DM annihilation process (or in the nuclear interactions from which the secondary fluxes are originated) without any sizable correlations among them, and then form the antinuclei if their relative momenta is smaller than $\pcoal$. In order to improve on this assumption and to include possible correlations or anticorrelations among the antinucleons (which would respectively increase or decrease the antinuclei production), or displaced production from heavier antibarions decay (which would then decrease antinuclei formation), coalescence has also been studied recently on an event-by-event basis with Monte Carlo generators. 
Although these analyses, which are typically based on the particle physics generators 
\textsc{Pythia} \cite{Kadastik:2009ts, Ibarra:2013qt, Fornengo:2013osa, Herms:2016vop}
or 
\textsc{Herwig} \cite{Dal:2014nda} are expected to provide
a detailed information on antinucleon production, one has to be aware of the fact that they can suffer from systematical uncertainties. 
These generators are built and trained to correctly reproduce the showering of elementary standard model particles into those
final states of a cascade that are actually looked for at accelerators: this does not guarantee that the individual properties of each particle in the cascade is properly modelled in the phase space specifically relevant for the formation of cosmic antinuclei. In fact, different Monte Carlo generators can lead to significantly different results in some energy range for the produced antinucleus \cite{Dal:2012my}. We therefore believe that the two approaches (the uncorrelated original model and the event-by-event Monte-Carlo model) are both viable scenarios at the current level of (still incomplete) understanding of the coalescence process that leads to the formation of cosmic antinuclei.
In the following, we will therefore exploit the analytic model of antimatter coalescence and compare it to a specific Monte Carlo model for the case of DM annihilation into a $\bar b b$ final state. The difference between analytic and Monte Carlo approach will therefore represent an estimate of the systematic uncertainty affecting the calculation.

Let us  now shortly summarize the results of the analytic coalescence model, first considering 
antideuteron production in the reaction $p+p\rightarrow\db +X$ and later extending it to different initial states and to antihelium production.
The natural way to state differential cross sections is the Lorentz invariant form 
$E_i \,\,d^3 \sigma_i/dk_i^3$, where $E_i$ is the particle's total energy and $\vk_i$ its momentum. 
The production cross section for antideuteron $\sigma_\db$ is given by 
\begin{eqnarray}
  \label{eqn::COAL_IV}
  E_\db \frac{d^3 \sigma_\db}{dk_\db^3} 
    = \frac{1}{\sigmaTot} \frac{m_D}{m_p m_n}  \frac{4\pi}{3}  \frac{\pcoal^3}{8}   
      E_\pb \frac{d^3 \sigma_\pb}{dk_\pb^3}  E_\nb \frac{d^3 \sigma_\nb}{dk_\nb^3}.
\end{eqnarray}
Here $\sigma_\pb$ and $\sigma_\nb$ are the cross sections for antiproton and antineutron production in a $pp$ collision, respectively.
We exploit the analytic parametrization of the invariant antiproton production cross section 
of Ref.~\cite{diMauro:2014_pbarCrossSectionParameterization} and take into account that antineutron production is enlarged by 30\% compared to antiproton. 
The prefactor in~\eqnref{COAL_IV} contains the particle masses  $m_i$, the coalescence momentum 
$\pcoal = |\vec \pcoal| = |\vec k_p - \vec k_n|$\footnote{Note that in the literature, some papers use a different definition of the coalescence momentum. In our case, the relative momentum between antiprotons and antineutrons has to be smaller that $\pcoal$: $\vec\Delta = |\vec k_p - \vec k_n| \leq \pcoal$. In other papers the condition is set on $2\vec\Delta$, which implies that $\pcoal\rightarrow \pcoal/2$ and the factor $1/8$ disappears from \eqnref{COAL_IV} \cite{Fornengo:2013osa}.}, 
and the total $pp$ cross section $\sigmaTot$.
Explicitly, the antiproton and antineutron production cross sections term is evaluated as:
\begin{eqnarray}
  \label{eqn::COAL_V}
  \frac{d^3 \sigma_\pb}{dk_\pb^3}   \frac{d^3 \sigma_\nb}{dk_\nb^3} 
    &=& \frac{1}{2} \left[  \frac{d^3 \sigma_\pb}{dk_\pb^3}\left(\sS, \vk_\pb\right)  
                            \frac{d^3 \sigma_\nb}{dk_\nb^3} \left(\sS-2E_\pb, \vk_\nb \right) \right. \quad \\ \nonumber
                            &&\left. \phantom{ \frac{d^3 \sigma_\pb}{dk_\pb^3}} + (\pb \leftrightarrow \nb)  \right],  
\end{eqnarray}
with $\vk_\pb = \vk_\nb = \vk_\db/2$. This expression takes into account that the production of an $n\nb$ pair after the $p\pb$ pair happens at an effectively smaller CM energy, and vice versa.
To obtain the cross sections for antideuterons produced by heavier nuclei in the initial state we re-scale the $pp$ cross section according to the mass number $A$ in the initial, and use a factor $A^{0.7}$ both for projectile and target nuclei. This is a first order approximation which is 
compatible with the results of Ref. ~\cite{Duperray:2003_pbarCrossSectionParameterizationForPA} for antiprotons.
In case of antiprotons in the initial state, we replace
the antiproton production cross section with the antiproton scattering cross section, since the reaction is $\pb+p\rightarrow\pb+X$.
The differential cross section with antiprotons is not measured: we therefore approximate it by using the parametrization of
Ref.~\cite{Anderson:1967zzc} for proton scattering. All considerations for antihelium ($^3\overline{\mathrm{He}}$) production are similar to the ones discussed for antideuteron. Here we only state the expression
equivalent to \eqnref{COAL_IV}:
\begin{eqnarray}
  \label{eqn::COAL_He}
  E_\Heb \frac{d^3 \sigma_\Heb}{dk_\Heb^3} 
    &=& \frac{m_\Heb}{m_p^2  m_n}  \left( \frac{1}{\sigmaTot}  \frac{4\pi}{3}  \frac{\pcoal^3}{8}  \right)^2 \\ \nonumber
      && \qquad \times E_\pb \frac{d^3 \sigma_\pb}{dk_\pb^3} E_\pb \frac{d^3 \sigma_\pb}{dk_\pb^3}  E_\nb \frac{d^3 \sigma_\nb}{dk_\nb^3}, 
\end{eqnarray}
where the antiproton and antineutron cross sections are evaluated at $\vk_\pb=\vk_\nb=\vk_\Heb/3$. The generalization of \eqnref{COAL_V} to antihelium is
\begin{eqnarray}
  \label{eqn::COAL_V_He}
  &&  \frac{d^3 \sigma_\pb}{dk_\pb^3}  \frac{d^3 \sigma_\pb}{dk_\pb^3}   \frac{d^3 \sigma_\nb}{dk_\nb^3} 
    = \frac{1}{3} \left[  \frac{d^3 \sigma_\pb}{dk_\pb^3} \left(\sS,              \vk_\pb\right)  \right. \\ \nonumber
       &&       \phantom{ \frac{d^3 \sigma_\pb}{dk_\pb^3}}
                          \frac{d^3 \sigma_\nb}{dk_\nb^3} \left(\sS-2E_\pb,       \vk_\nb\right) 
                          \frac{d^3 \sigma_\pb}{dk_\pb^3} \left(\sS-2E_\pb-2E_\nb,\vk_\pb\right)   \quad \\ \nonumber
       &&\left. \phantom{ \frac{d^3 \sigma_\pb}{dk_\pb^3}} + (\pb \leftrightarrow \nb \, \pb) 
                            + (\pb \, \nb \leftrightarrow \pb)  \right]. 
\end{eqnarray}

Before deriving the antimatter source terms in the next section we recast \eqnref{COAL_IV} and \eqnref{COAL_He} into a suitable form for DM annihilation. If we assume spherical symmetry for all differential cross sections we can replace the Lorentz invariant form with the energy spectrum:
\begin{eqnarray}
  \label{eqn::XS_DM_transform_I}
  E\frac{d^3 N}{dk^3} = \frac{E}{4\pi k^2} \frac{dE}{dk} \frac{dN}{dE} = \frac{1}{4\pi k} \frac{dN}{dE}.
\end{eqnarray}
Applying \eqnref{XS_DM_transform_I} we get:
\begin{eqnarray}
  \label{eqn::XS_DM_transform_D}
  \frac{d N_\db}{dE_\db} 
    =  \frac{m_D}{m_p m_n} \frac{4}{3} \frac{\pcoal^3}{8 k_\db}
       \frac{d N_\pb}{dE_\pb}   \frac{d N_\nb}{dE_\nb}. 
\end{eqnarray}
and
\begin{eqnarray}
  \label{eqn::XS_DM_transform_II_He}
  \frac{d N_\Heb}{dE_\Heb} 
    =  \frac{m_\mathrm{He}}{m_p^2 m_n } 3 \left( \frac{\pcoal^3}{8 k_\Heb} \right)^2
       \frac{d N_\pb}{dE_\pb} \frac{d N_\pb}{dE_\pb}  \frac{d N_\nb}{dE_\nb}. 
\end{eqnarray}
for antideuteron and antihelium, respectively. In principle, these equations are also affected by the energy subtraction  shown in \eqnref{COAL_V} and \eqref{eqn::COAL_V_He}, but since the DM mass is far above the production threshold the effect on the spectrum is below 1\%.
We use the energy spectra for antiprotons which are publicly available in~\cite{Cirelli:2010xx}, and we take antineutron and antiproton spectra to be equal. The energy spectra include electro-weak correction, namely radiation of $W$ or $Z$ bosons from the 
standard model final states.

\subsection{\label{sec::source}Source term determination}

The source term of CR antimatter in our Galaxy contains the standard astrophysical term and potentially a DM contribution. 
We first consider the antimatter produced by cosmic ray spallation, which forms the background for any DM search. 
This term is produced in the interaction of a CR species $i$ on the interstellar medium component $j$. 
The source term $q_{ij}$ for antideuterons is therefore given by a convolution of the CR flux $\phi_i$ and the ISM density $n_{\mathrm{ISM},j}$
with the production cross section $\sigma_{ij}$:
\begin{eqnarray}
\label{eqn::sourceTerm_1}
  q_{ij}(E_\db) &=& \int\limits_{E_{\rm th}}^\infty dE_i \,\, 4\pi \,n_{\mathrm{ISM},j} \, 
                      \phi_i  (E_i) \, \frac{d\sigma_{ij}}{d E_\db}(E_i , E_\db).
\end{eqnarray}
Here $E_{\rm th}$ is the energy threshold for antideuteron production. 
It is interesting to note that the production threshold for antideuterons in $pp$ collisions in the 
center-of-mass frame is $\sS \geq 6 m_p$ and consenquenty $E = (s-2m_p^2)/(2m_p) \geq 17 m_p$ in the laboratory frame, where one of the protons is at rest (and which corresponds to the process occurring in the Galaxy).
This threshold is much larger than for antiproton production ($E \geq  7 m_p$) and represents the main reason 
for the good signal-to-background ratio in DM searches with antideuterons \cite{Donato:1999gy}.
The ISM consists of hydrogen and helium, and we assume a constant density of 1 $\mathrm{cm^{-3}}$ and 0.1 $\mathrm{cm^{-3}}$ in the Galactic disk, respectively.
The fluxes of proton, helium, and antiprotons are inferred directly from AMS-02 data
\cite{AMS-02_Aguilar:2015_ProtonFlux, AMS-02_Aguilar:2015_HeliumFlux, AMS-02_Aguilar:2016_AntiprotonFlux}
which have been demodulated from solar modulation in the force-field approximation \cite{Fisk:1976_SolarModulation}, by taking a solar modulation potential of 600~MV. Note that the effect of solar modulation mostly affects the low-energy tail of the 
projectile spectrum which does not contribute to the determination of the secondary antideuterons, due to the large energy threshold for production.
More details on the source term calculations may be found in \cite{Donato:2017ywo}. In summary, antideuterons are dominantly produced by CR proton and helium, while a small contribution arises from antiprotons.

The DM source term originates from annihilation of two DM particles $\chi$ into standard model particles. For definiteness, let us concentrate on a pure $b$-quark final state: $\chi \chi \rightarrow b \bar{b}$. The source term is given by:
\begin{eqnarray}
\label{eqn::sourceTerm_DM}
  q_\mathrm{DM}(E_\db, \vec{x}) &=& 
      \frac{1}{2} \left( \frac{\rho(\vec{x})}{m_\mathrm{DM}}\right)^2  \sv_{b\bar{b}}  \frac{d N_\db^{b\bar{b}}}{d E_\db} ,
\end{eqnarray}
where $\rho$ is the spatial-dependent DM mass density, $\sv_{b\bar{b}}$ is the thermally averaged rate for annihilation into a $\bar b b$ quark pair, and $d N_\db^{b\bar{b}}/d E_\db$ is the antideuteron energy spectrum per annihilation event.
The factor $1/2$ corresponds to a self-conjugate DM particle forming the mass of the DM halo (being 1/4 for a non self-conjugate DM). 
We assume a NFW DM density profile~\cite{Navarro:1995iw}: 
\begin{eqnarray}
\label{eqn::NFW}
  \rho_{\mathrm{NFW}}(r) &=&  \frac{\rho_h}{(r/r_h)(1+r/r_h)^2},
\end{eqnarray}
with a scale radius $r_h=20$~kpc and a halo density $\rho_h$ normalized such that the local DM density is 0.43~$\mathrm{GeV/cm^3}$ \cite{Salucci:2010qr} at the position of the Sun $r =r_\odot = 8$~kpc. 
The same formulas above are valid for also antihelium, with $d N_\db^{b\bar{b}}/d E_\db$ replaced by $d N_\Heb^{b\bar{b}}/d E_\Heb$.

Changing to a different DM density profile affects the results only mildly since CRs mostly probe a relatively local portion of the galactic DM, as was shown in \cite{Donato:2003xg, 2003A&A...404..949M}. We have nevertheless explicitly calculated the difference between the NFW profile and a cored Burkert profile (5~kpc scale radius) to be about 30\%. Moreover, this effect is degenerate with $\sv$: if the Burkert profile decreases the antiproton signal, the fit in CuKrKo requires a larger value of $\sv$. Overall, the estimate for an antideuteron or antihelium signal is therefore unchanged.
\begin{figure}[b!]
	{\includegraphics[width=0.5\textwidth]{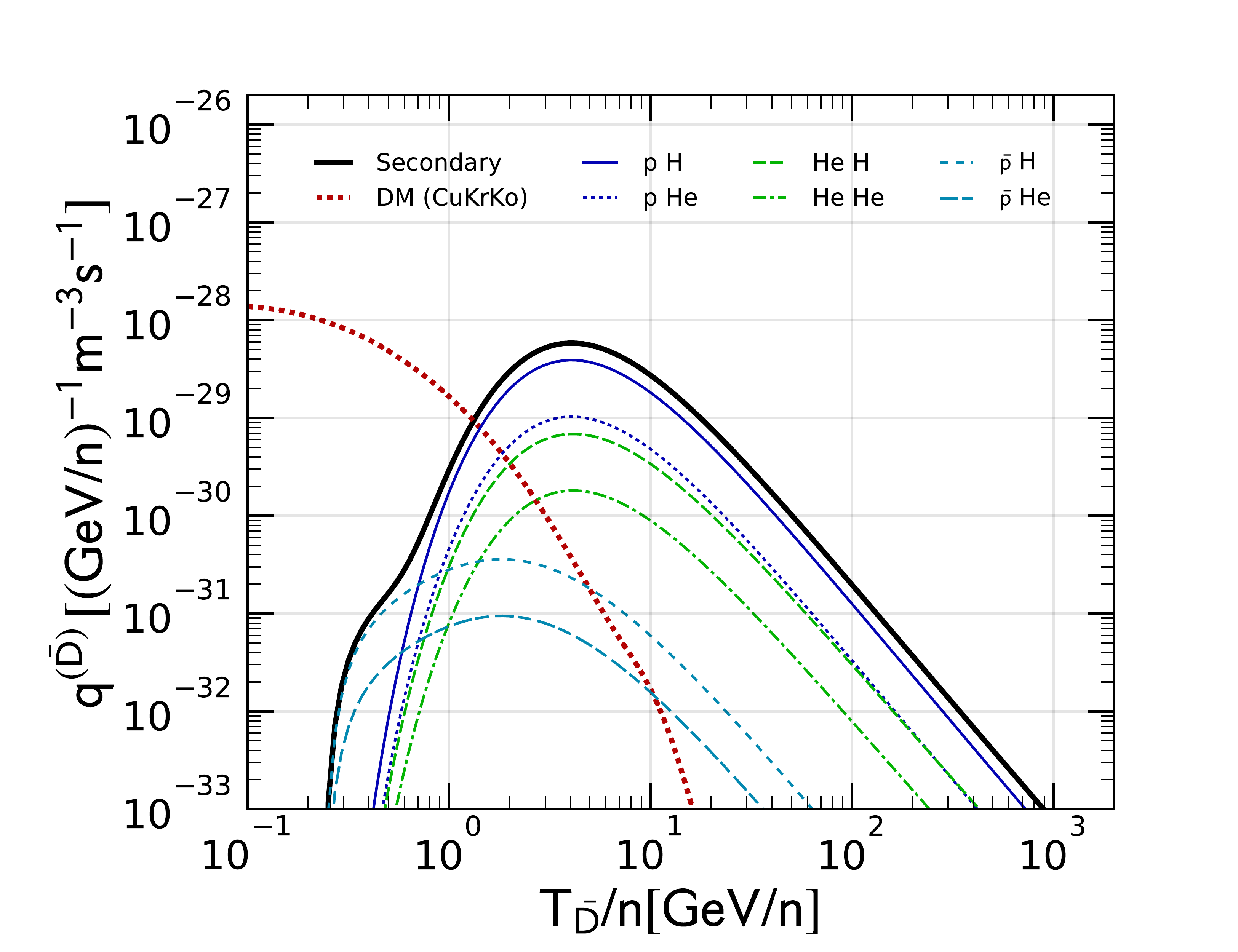} } \\
	{\includegraphics[width=0.5\textwidth]{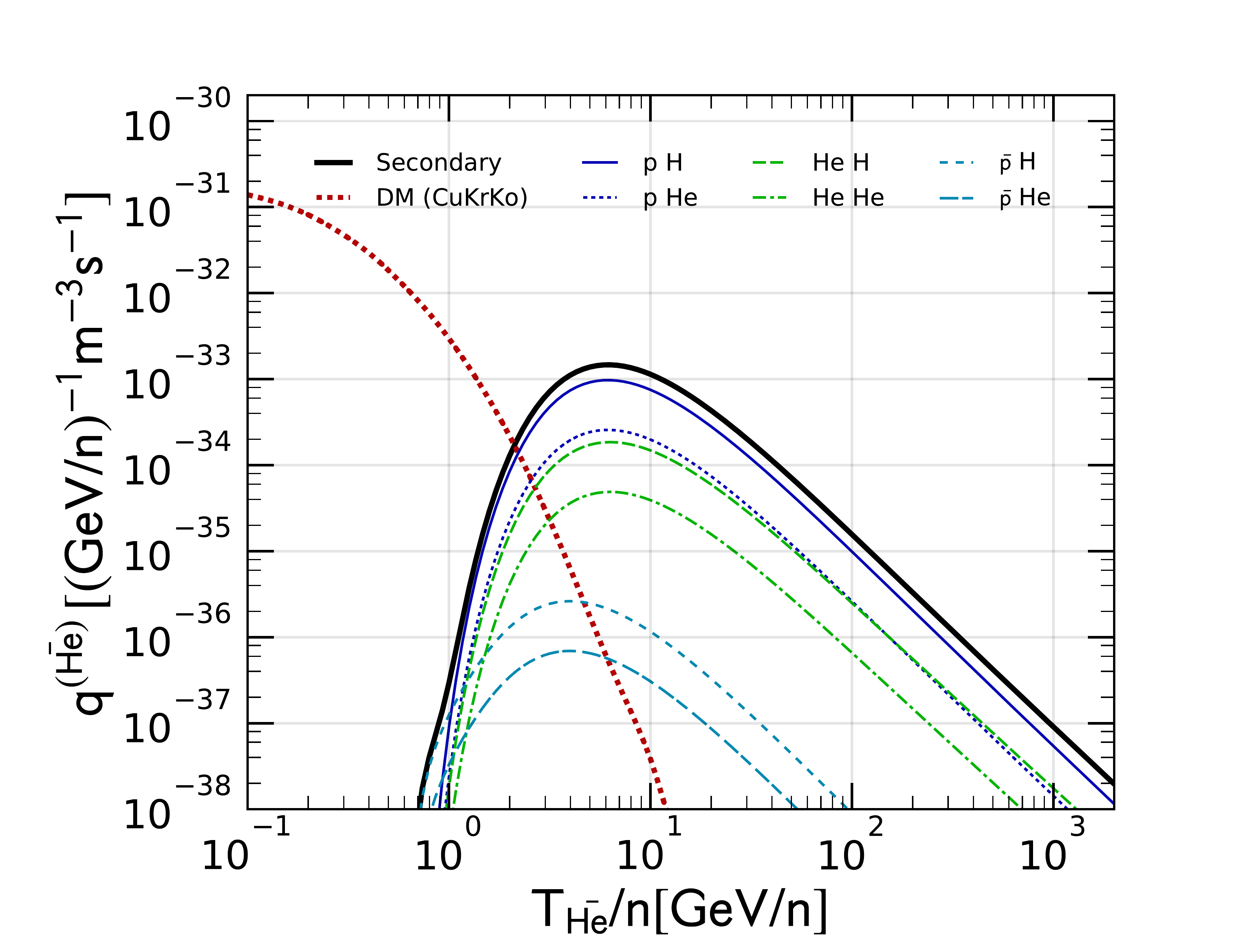}} 
	\caption{ Local source term for the ISM secondary  and DM primary  antideuteron (upper panel) 
	          and antihelium (lower panel).
	          The secondary term is also shown in its single components given by cosmic p, He and $\bar{p}$ interacting with the ISM. 
	          The DM signal corresponds to the best fit of the antiproton excess in CuKrKo for annihilation into $b\bar{b}$, mass
	           $\mDM=71$~GeV, annihilation rate $\sv = 2.6 \cdot10^{-26}\,\mathrm{cm^3/s}$, 
	          and a local DM density of 0.43~$\mathrm{GeV/cm^3}$. We use  a coalescence momentum of {$\pcoal = 160$~MeV}. 
	          }	
  \label{fig::antimatter_sourceterm}
\end{figure}

The coalescence process for secondary and DM antimatter involves significantly different kinematics. While the DM annihilation takes place at rest, the secondary production through CRs is highly boosted. Moreover, DM annihilation involves the interaction of non-nuclear species, while the secondary production is a nuclear process. This implies that in general the coalescence momenta of the non-nuclear and nuclear processes {\sl might} not be the same. 
For DM annihilation, a tuning for the coalescence momentum is usually derived from the measured antideuteron production from the $Z$-boson decay in the ALEPH experiment \cite{Schael:2006fd}. Since the initial state is not hadronic, this setup can be considered to be closer to the situation of DM annihilation. The value derived for the coalescence momentum by adopting the non-correlated coalescence is {$\pcoal= (160 \pm 19) $~MeV} \cite{Fornengo:2013osa}.
Antimatter production in $pp$ collision instead might be affected by QCD corrections in the initial state and give different values for $\pcoal$. 
Very recently the ALICE experiment measured the production of antideuteron and antihelium in $pp$ collisions at three different 
$\sqrt{s}$: 0.9, 2.76, and 7~TeV \cite{Acharya:2017fvb} and provided the so called $B_2$ and $B_3$ parameters defined as:
\begin{eqnarray}
  \label{eqn::COAL_BA}
  E_A \frac{d^3 N_A}{dk_A^3} 
    &=& B_A   \left( E_\pb \frac{d^3 N_\pb}{dk_\pb^3} \right)^Z \left( E_\nb \frac{d^3 N_\nb}{dk_\nb^3} \right)^{A-Z}.
\end{eqnarray}
A comparison to \eqnref{COAL_IV} and \eqnref{COAL_He} reveals a relation between these parameters and $\pcoal$:
\begin{eqnarray}
  \label{eqn::COAL_B2_B3}
  B_2= \frac{m_D          }{m_p   m_n}        \frac{\pi \pcoal^3}{6 }            \quad \text{and} \quad
  B_3= \frac{m_\mathrm{He}}{m_p^2 m_n} \left( \frac{\pi}{6} \pcoal^3 \right)^2.
\end{eqnarray}
ALICE provides $B_2$ as function of transverse momentum from $\pT/A=0.4$ to 1.5~GeV. Noting that the cosmic-rays source term calculation enhances the 
low $\pT$ values, we estimate $B_2$ to be between 0.01 and 0.02~$\mathrm{GeV^2}$ which implies 
a coalescence momentum between 208 and 262~MeV. On the other hand, $B_3$ is only measured at $\sqrt{s}=7$~TeV and converges between $1\cdot10^{-4}$ 
and $3\cdot10^{-4}$~$\mathrm{GeV^4}$. Interestingly, although the coalescence into antihelium could in principle be different from the antideuteron case, 
it leads to similar coalescence momenta between 218 and 261~MeV. 
The recent ALICE measurements therefore hint to a larger coalescence momentum (see also~\cite{Blum:2017qnn}), similar for antideuteron and antihelium. In order to somehow bracket the uncertainty on this parameter, we provide in \secref{results} an explicit comparison between the two scenarios with a lower (160 MeV) and higher (248 MeV) value of $\pcoal $.   
Let us notice that the coalescence momentum $\pcoal$ could also change with the energy at which the process of antinuclei formation occurs \cite{Fornengo:2013osa}. This implies that the value of $\pcoal$ determined from the high-energy ALICE data might not be adequate for the cosmic rays energies relevant for the cosmic antinuclei production. However, from the investigation of antideuteron production at different energies \cite{Aramaki:2015pii,Blum:2017qnn,Acharya:2017fvb}, no clear evolution is seen the  value of $\pcoal$. For this reason, we here assume that $\pcoal$ is independent of energy and we adopt the value obtained from ALICE data throughout in the determination of the secondary components. What is instead observed in the ALICE data is a dependence on the transverse momentum $p_T$ of the process \cite{Acharya:2017fvb}: in this case, we adopt the $\pcoal$ value derived for low-$p_T$, since this is the kinematical regime which dominates in the source term integral in \eqnref{sourceTerm_1}. For alternative approaches to the determination of $\pcoal$ see \cite{Ibarra:2012cc,Dal:2014nda,Dal:2015sha}.

To conclude this section, we show in
\figref{antimatter_sourceterm} the antideuteron and antihelium {\sl source terms} of astrophysical secondaries and the potential 
indication for DM from CuKrKo in the $\bar b b$ channel. The separate contributions to the astrophysical term from various initial states are also shown. 
As expected, at low kinetic energies the DM production largely dominates, due to the kinematical cut offs discussed above.
Note that these are the {\sl local} source terms. On the scale of our Galaxy they exhibit different spatial distribution which, leads to an enhancement of the DM {\sl global} source term as we discuss in the next paragraph. 

\subsection{\label{sec::prop}Propagation in the Galaxy}
\begin{table}[b!]
\caption{Summary of the propagation parameters.}
\label{tab::propagation}
\begin{tabular}{lccc}
\hline \hline
 Parameter                    & CuKrKo  & MED      & MAX     \\ \hline 
 $D_0$~$\mathrm{[10^{28}\;cm^2/s}]$ & 9.8 &        &         \\
 $K_0$~$\mathrm{[kpc^2/Myr}]$ &         &  0.0112  & 0.0765  \\
 $\delta$                     &  0.25   &  0.70    &   0.46  \\
 $V_c$~[km/s]                 &  45     &  12      &   5     \\
 $V_A$~[km/s]                 &  29     &          &         \\
 $L$~[kpc]                    &  5.4    &   4      &  15     \\ \hline \hline
\end{tabular}
\end{table}

The propagation of CRs in the Milky Way is described by a diffusion equation.
We use the \textsc{Galprop} code \cite{Strong:1998fr} with the same set-up from 
CuKrKo to solve this equation numerically. The diffusion halo in the Galaxy is
modelled assuming cylindrical symmetry with a radial extension of 20~kpc and a halo half-height $L$. We adopt isotropic diffusion with a normalization $D_0$ and slope $\delta$ of the diffusion coefficient. We include reacceleration by Alfven magnetic waves of velocity $V_A$, convection with a constant velocity $V_c$, and continuous energy losses. DM annihilation is described with two parameters $m_\DM$ and $\sv$. As a default, we add the propagation parameters of the best fit with DM from CuKrKo which are summarized in \tabref{propagation}. More details are given in CuKrKo. 
Note that the analysis of CuKrKo is performed only on proton, helium, and antiproton data but there is an analysis with a similar indication for DM \cite{Cui:2016ppb} which is tuned to AMS-02 B/C data \cite{Aguilar:2016vqr}. 

In its standard configuration \textsc{Galprop} does not calculate antideuteron (and antihelium) fluxes, therefore we upgrade the code accordingly. We use the production cross section for secondary and DM antideuteron and antihelium as derived in the previous section. Furthermore, we include tertiary antideuterons, which are inelastically scattered secondaries.
In other words, 
antideuterons might scatter on the ISM without annihilating but loosing a significant fraction of their original energy. This scattered antideuteron can be treated as a new component of the source term, conventionally called {\it tertiary}. The tertiary component is calculated in analogy to the secondary source term with \eqnref{sourceTerm_1}, where we use the propagated secondary antideuteron flux in the integral and replace the production cross section with the non-annihilation cross section of the reaction 
$\db + p \rightarrow \db + X $. 
To estimate this cross section we proceed as in~\cite{Duperray:2005si}. We use the scarce data of 
antideuteron scattering accompanied by pion production from~\cite{Baldini:1988} to determine the size of the absolute cross section. Then we apply the differential form of proton scattering from~\cite{Anderson:1967zzc}. For helium in the initial state we multiply by a factor $4^{0.8}$. 
In principle there is a corresponding loss term in the secondary antideuteron flux, but this is a negligible effect at the one percent level.
For antihelium scattering we assume a non-annihilation cross section increaded by a factor $3/2$.
Notice that there is a similar process for DM antideuterons. We calculate these \textit{secondary-DM} antideuteron finding that their contribution is suppressed by two orders of magnitude, with respect to the primary DM signal. Nonetheless, they are included in the DM fluxes shown in the following.

We compare our results against the analytic diffusion model of~\cite{Maurin:2002ua}.
For the specific case of antideuteron, this was already discussed in~\cite{Donato:2008yx} and for antihelium in~\cite{Cirelli:2014qia}.
We shortly recall the model and specify the relevant physical quantities. 
Propagation of nuclei in the Galaxy is described by the diffusion equation:
\begin{eqnarray}
\label{eqn::DiffusionEq}
  \frac{\partial f}{\partial t} &-& K(E) \cdot \nabla^2 f + \partial_z \left( \mathrm{sign}(z) f V_\mathrm{conv} \right)  \\ \nonumber
                                &=& Q - 2 h \delta(z) \Gamma_\mathrm{ann}f, 
\end{eqnarray}
where $f=d N/dE$ is the energy derivative of the number density $N$. 
The single terms describe: diffusion, modeled with a diffusion coefficient $K(E)= K_0 \beta (R/(GV))^{\delta}$; convection, with a velocity $V_\mathrm{conv}$; a source term $Q$; annihilation losses, described through the annihilation rate $\Gamma_\mathrm{ann}$. Furthermore, $K_0$ is a normalization and $\delta$ the spectral index of the diffusion constant, $\beta$ is velocity in units of speed of light and $R$ is the particle rigidity. 
\eqnref{DiffusionEq} is solved analytically in terms of Bessel functions. The solution at the solar position in the Galaxy is given
by~\cite{Maurin:2002ua}:
\begin{eqnarray}
\label{eqn::DiffusionEq_solution}
  f &=& \sum\limits_{n=1}^\infty J_0 \left( \xi_n \frac{r_\odot}{R} \right) 
                                                      \exp \left( -\frac{V_\mathrm{conv} L}{2K(E)} \right) \\ \nonumber
                                                      && \hspace{1.5cm} \times \frac{y_n(L)}{A_n \sinh(S_n L /2)}
\end{eqnarray}
where:
\begin{eqnarray}
\label{eqn::DiffusionEq_solution_II}
  Q_n &=& \frac{4}{J_1^2(\xi_n) R^2} \int\limits_0^R dr \, r J_0 (\xi_n r /R) Q(r,z),  \\ \nonumber 
  y_n &=& \int\limits_0^Z dz \, \exp \left( \frac{V_\mathrm{conv} (Z-z)}{2K} \right) \\ \nonumber
                   && \hspace{1cm} \times \sinh(S_n(Z-z)/2) Q_n,  \\ \nonumber 
  A_n &=& 2h\Gamma_\mathrm{ann} + V_\mathrm{conv} + K S_n \coth(S_n L /2),  \\ \nonumber 
  S_n &=& \left( V_\mathrm{conv}^2/ K^2 + 4 \xi_n^2/R^2) \right)^{1/2},  \\ \nonumber 
  \Gamma_\mathrm{ann} &=& \left( n_\mathrm{H} + 4^{2/3}n_\mathrm{He} \right) v \sigma_\mathrm{ann}. \\ \nonumber 
\end{eqnarray}
Here $J_0$ and $J_1$ are the zero- and first-order Bessel functions, while $\xi_n$ is the $n$-th zero of $J_0$. 
The spatial variables $r$ and $z$ are radial distance and height above the Galactic plane in cylindrical coordinates, respectively.
$L$ is the diffusive halo half-height, $h=0.1$~kpc the half-height of the galactic disk, and 
$R=20$~kpc is the radial extension of the Galaxy.
The annihilation cross section $\sigma_\mathrm{ann}$ is the difference between the total $\sigmaTot$ and the elastic $\sigmaEl$ cross sections.
There is only a measurement of the total deuteron-antiproton cross section
\cite{Agashe:2014kda}, which by symmetry is equal to antideuteron-proton.  
So, to infer the antideuteron annihilation cross section we approximate it from the $p\pb$ scattering as:
\begin{eqnarray}
\label{eqn::DiffusionEq_solution_III}
  \sigma_\mathrm{ann}^{\db p} \approx \frac{\sigmaTot^{\db p}}{\sigmaTot^{\pb p}} (\sigmaTot^{\pb p} - \sigmaEl^{\pb p}). 
\end{eqnarray}
Elastic and total $p\pb$ cross section data are taken from~\cite{Agashe:2014kda}. 
For antihelium we re-scale according to the mass number 
$\sigma_\mathrm{ann}^{\Heb p} = 3/2 \sigma_\mathrm{ann}^{\db p}$. 
Finally, the CR flux is related to $f$ by:
\begin{eqnarray}
\label{eqn::DiffusionEq_solution_IV}
  \phi &=& \frac{\beta c}{4\pi} f.  
\end{eqnarray}
More details about propagation and the analytic solution of \eqnref{DiffusionEq_solution} 
are given in \cite{Maurin:2002ua} and references therein.

The free propagation parameters are fixed to two benchmark scenarios. 
The analysis of Ref. \cite{Donato:2003xg} identified three different parameter sets (MIN, MED, and MAX) which were consistent 
with B/C measurements and corresponded to significant variations of the amount of antiprotons from DM annihilation. 
Since the MIN scenario seems strongly disfavored  by recent analyses on new AMS data \cite{DiMauro:2014iia,Lavalle:2014kca,Genolini:2015cta,Boudaud:2016jvj}, we adopt here the propagation benchmarks MED and MAX, whose   propagation parameters are summarized in \tabref{propagation}.

\begin{figure*}[t!]
  {\includegraphics[width=0.5\textwidth]{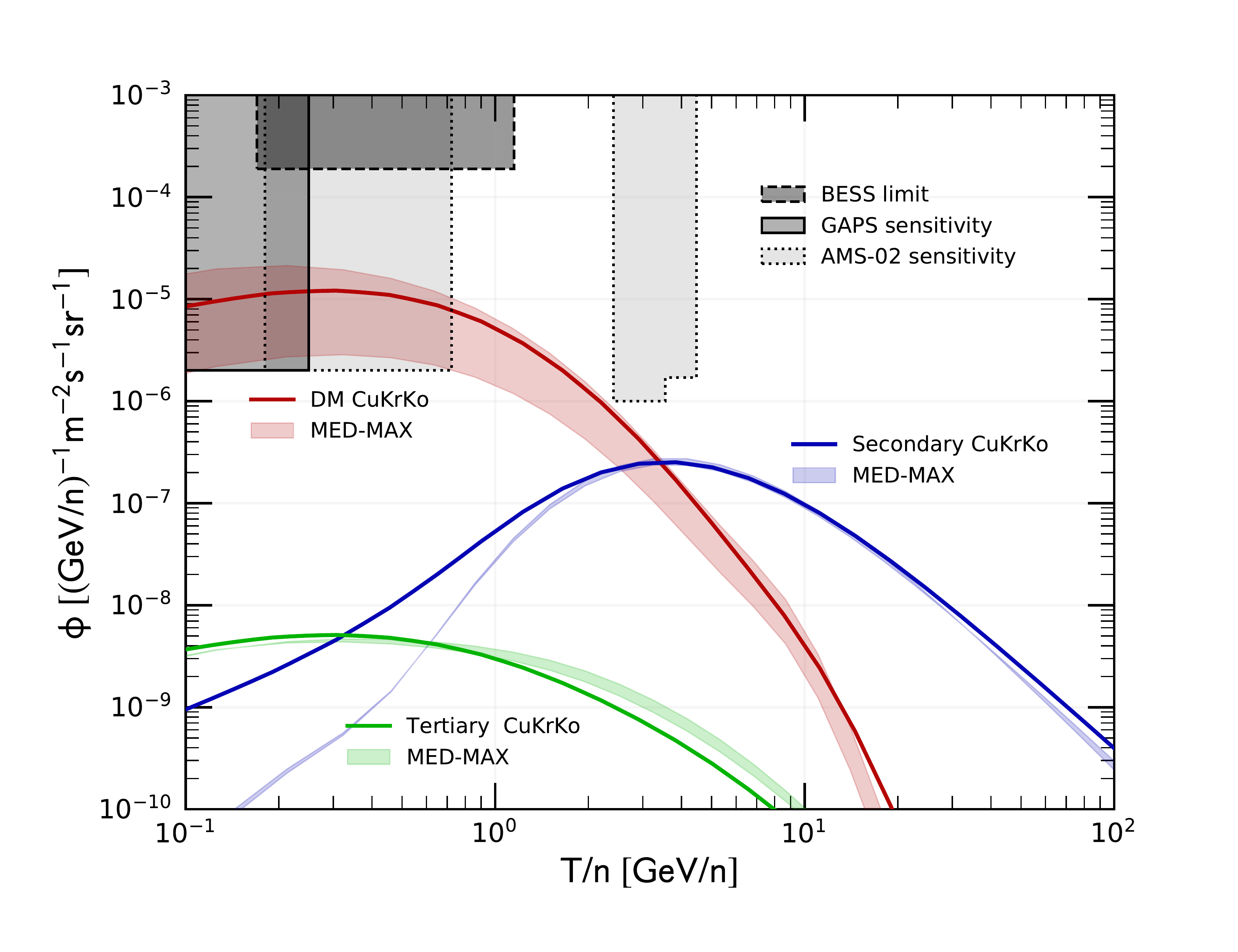}}{\includegraphics[width=0.5\textwidth]{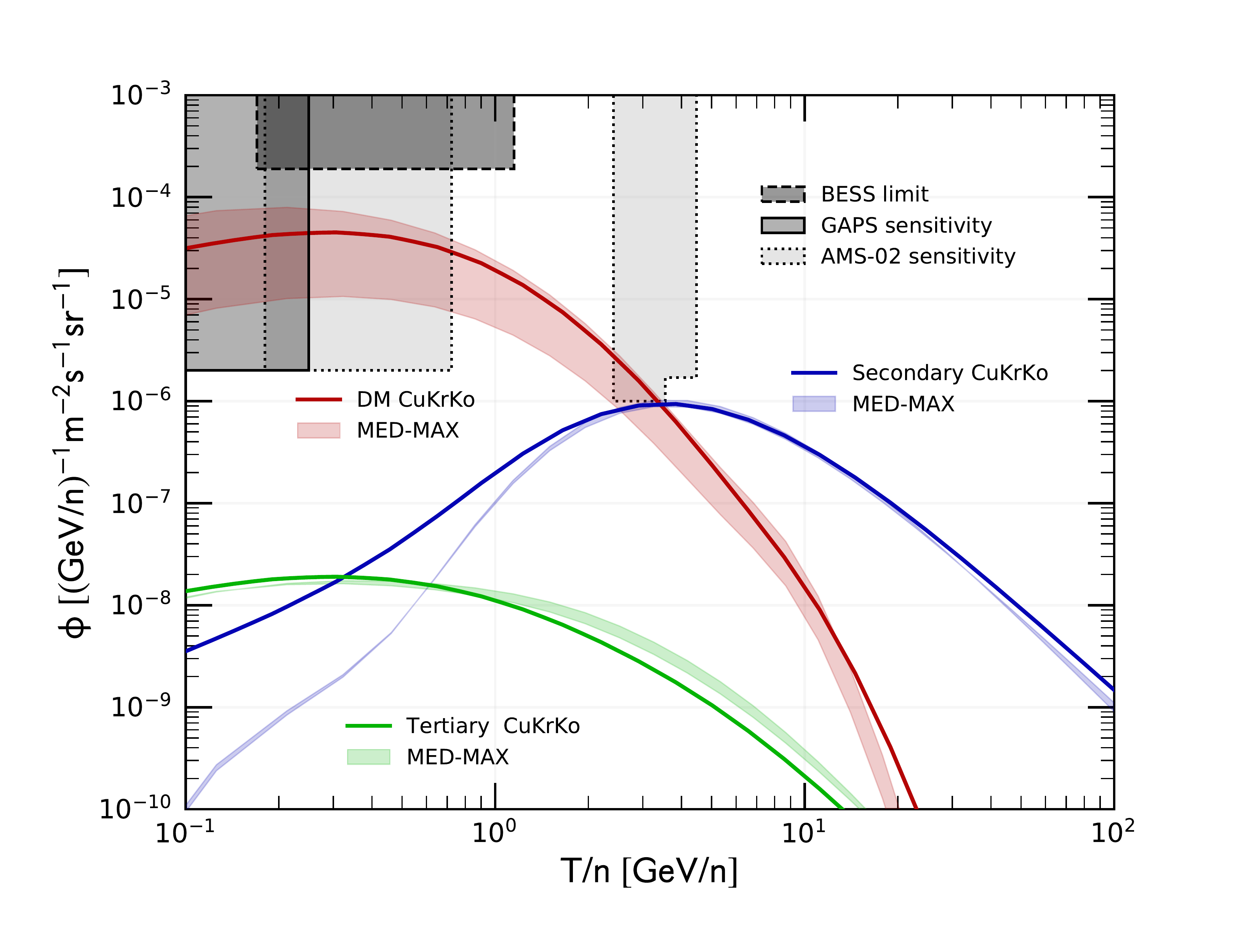}} 	
	\caption{ Antideuteron flux for  secondaries in the ISM and the potential DM signal, corresponding 
            to generic $b\bar{b}$ annihilation 
            from the excess in CuKrKo. We show the different propagation models MED and MAX, which are constrained to fit 
            B/C data in Ref. \cite{Donato:2003xg}.
            CuKrKo corresponds to the propagation parameters obtained from the best fit of $b\bar{b}$ DM in \cite{Cuoco:2016eej}.
            All fluxes are derived in the analytic coalescence model with {$\pcoal=160$~GeV}~(left panel) 
            and $\pcoal=248$~GeV~(right panel). 
            Solar modulation is treated in the force-field approximation with a potential of $\phi=400$~MV.
            Additionally, the current limit by the BESS experiment (95\%~CL) \cite{Fuke:2005it}, 
            the AMS-02 sensitivity of \cite{Aramaki:2015pii},
            and the expected sensitivity for GAPS (99\%~CL) \cite{Aramaki:2015laa} are displayed. }
	\label{fig::antideuteron_flux}
\end{figure*}

\begin{figure*}[t!]
	\subfloat[Coalescence model]{\includegraphics[width=0.5\textwidth]{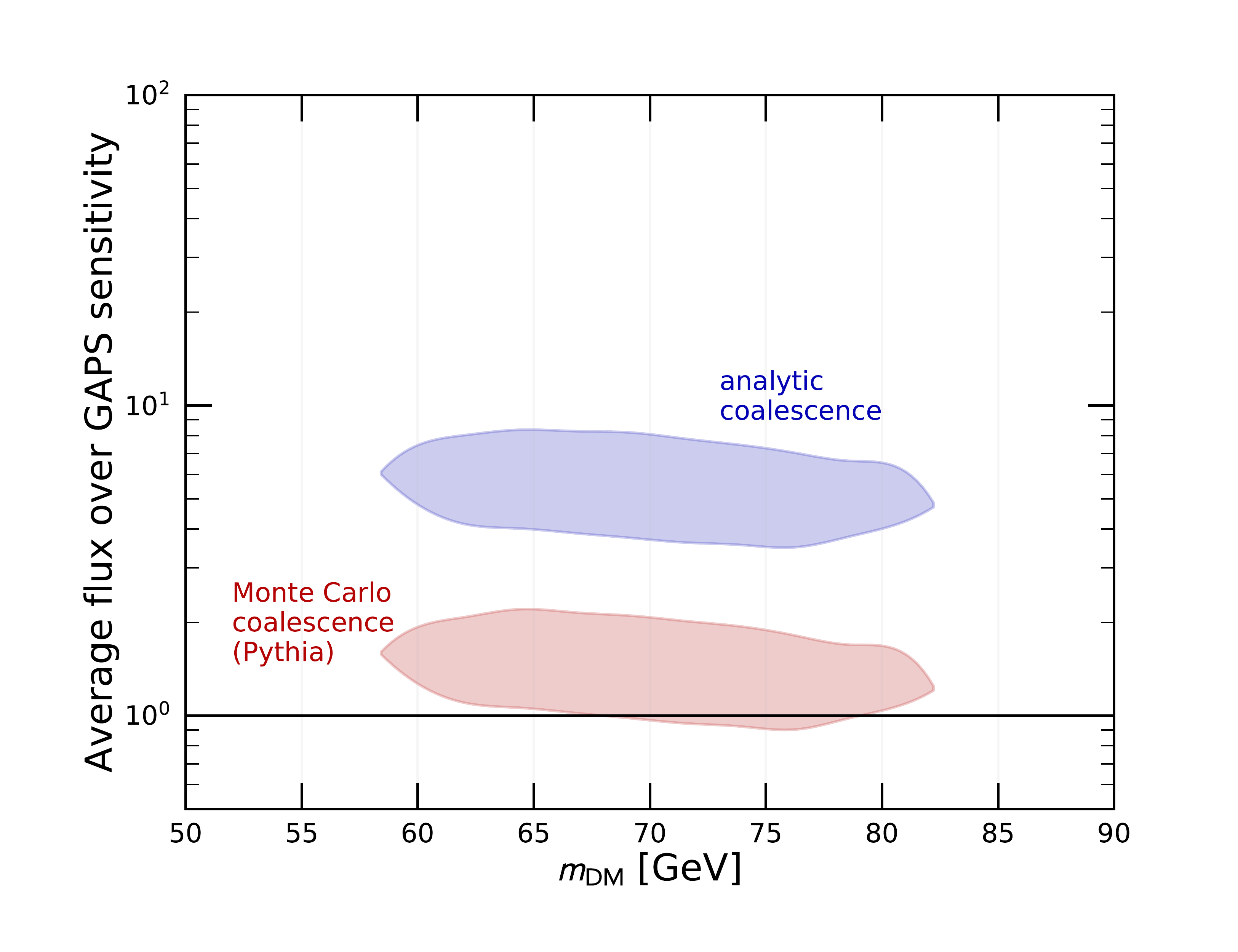}}\subfloat[Coalescence momentum]{\includegraphics[width=0.5\textwidth]{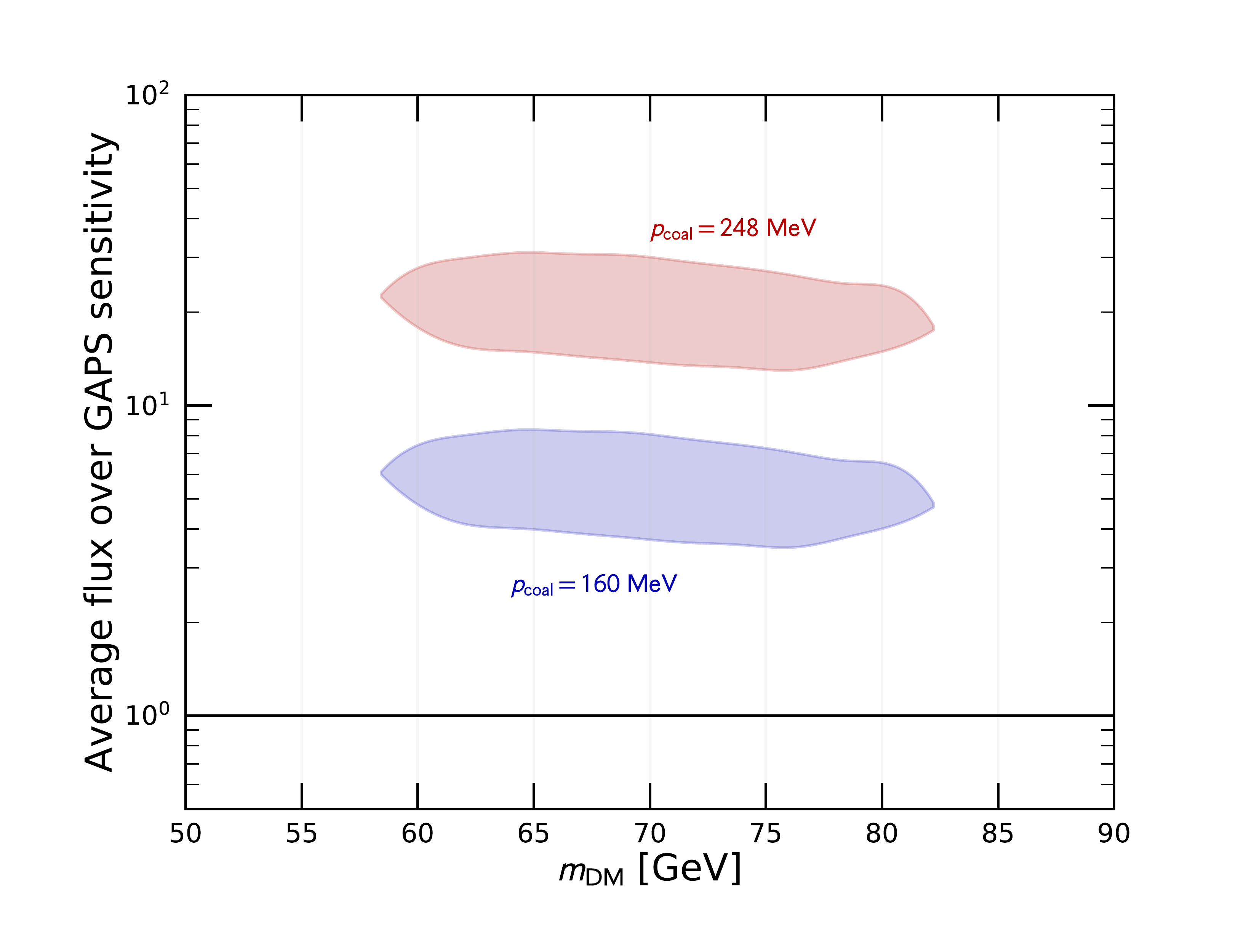}} \\
	\subfloat[Solar modulation]{\includegraphics[width=0.5\textwidth]{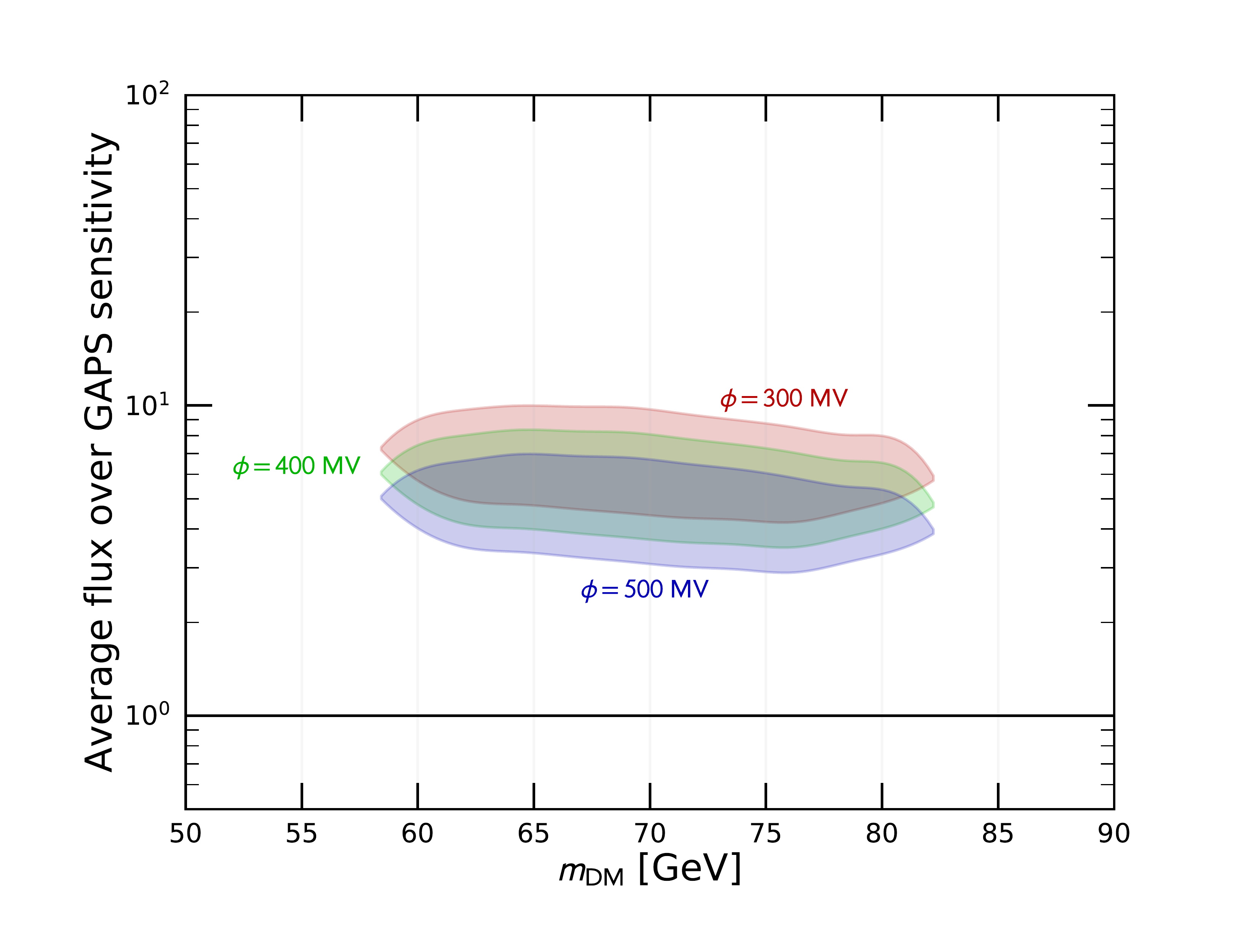}}\subfloat[Propagation model]{\includegraphics[width=0.5\textwidth]{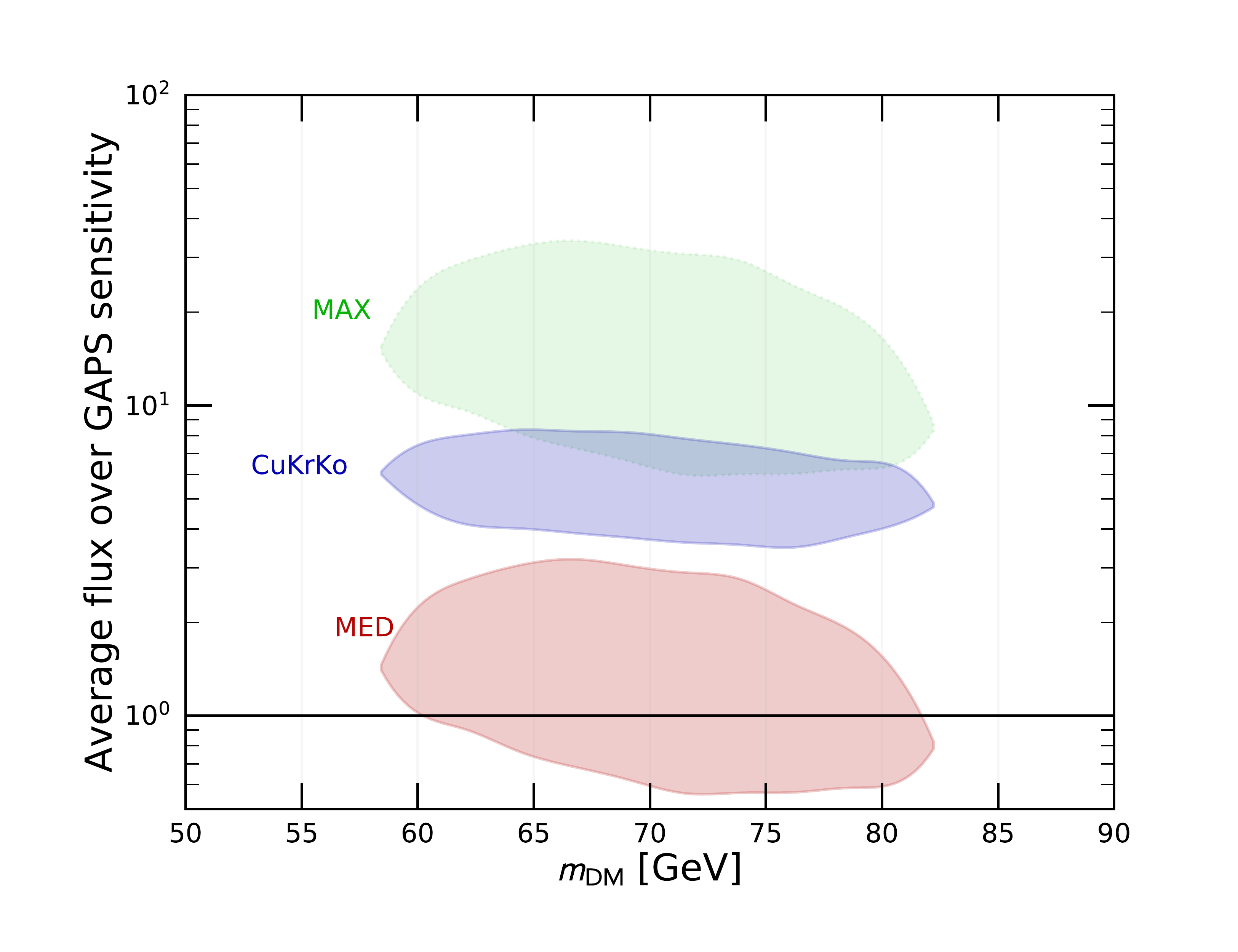}} 
  \caption{  Average antideuteron flux in the GAPS energy range divided by the expected GAPS sensitivity
             of $2.0 \times 10^{-6}$~$\mathrm{m^{-2}s^{-1}sr^{-1}(GeV/n)^{-1}}$ \cite{Aramaki:2015laa}. 
             The areas correspond to the 2$\sigma$ contours from the DM hint properties in CuKrKo.
             The reference case (blue contour) relies on the analytic coalescence model, 
             with a coalescence momentum of {$p_\coal = 160$~MeV}, solar modulation
             in the force-field approximation with a potential of $\phi=400$~MV, and the propagation parameters taken 
             (individually for each point in the contour) from CuKrKo. We compare against a Monte Carlo based
             coalescence from~\cite{Fornengo:2013osa} in panel\;(a), a larger coalescence momentum 
             as might be justified by~\cite{Acharya:2017fvb} in panel\;(b), a
             different solar modulation in panel\;(c), and different propagation parameters in panel\;(d).
             The MAX contour should be treated with caution since its propagation parameters are probably 
             in conflict with the DM signal of CuKrKo. We show the contour for the sake of completeness.
           }
	\label{fig::antideuteron_uncertatinty}
\end{figure*}

\begin{figure}[t!]
	\includegraphics[width=.5\textwidth]{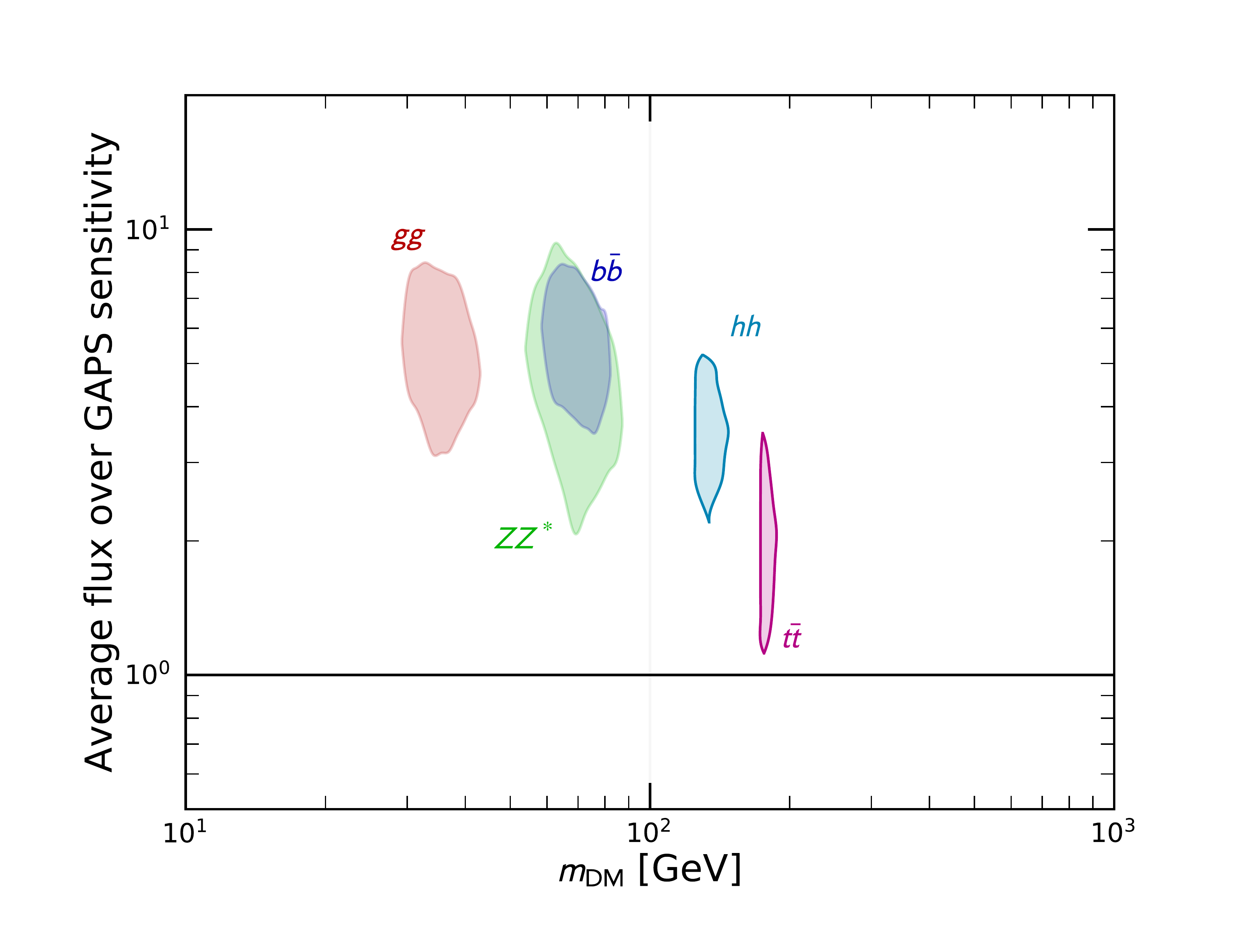} 
  \caption{  Same as \figref{antideuteron_uncertatinty}. The reference case corresponding to generic DM annihilation into $b\bar{b}$ final states is shown along with other standard model final states $gg$, $ZZ^*$, $hh$, and $t\bar{t}$. 
             The 2$\sigma$ countours are taken from~\cite{Cuoco:2017rxb}.
          }
	\label{fig::antideuteron_final_states}
\end{figure}

\begin{figure*}[t!]
	  {\includegraphics[width=0.5\textwidth]{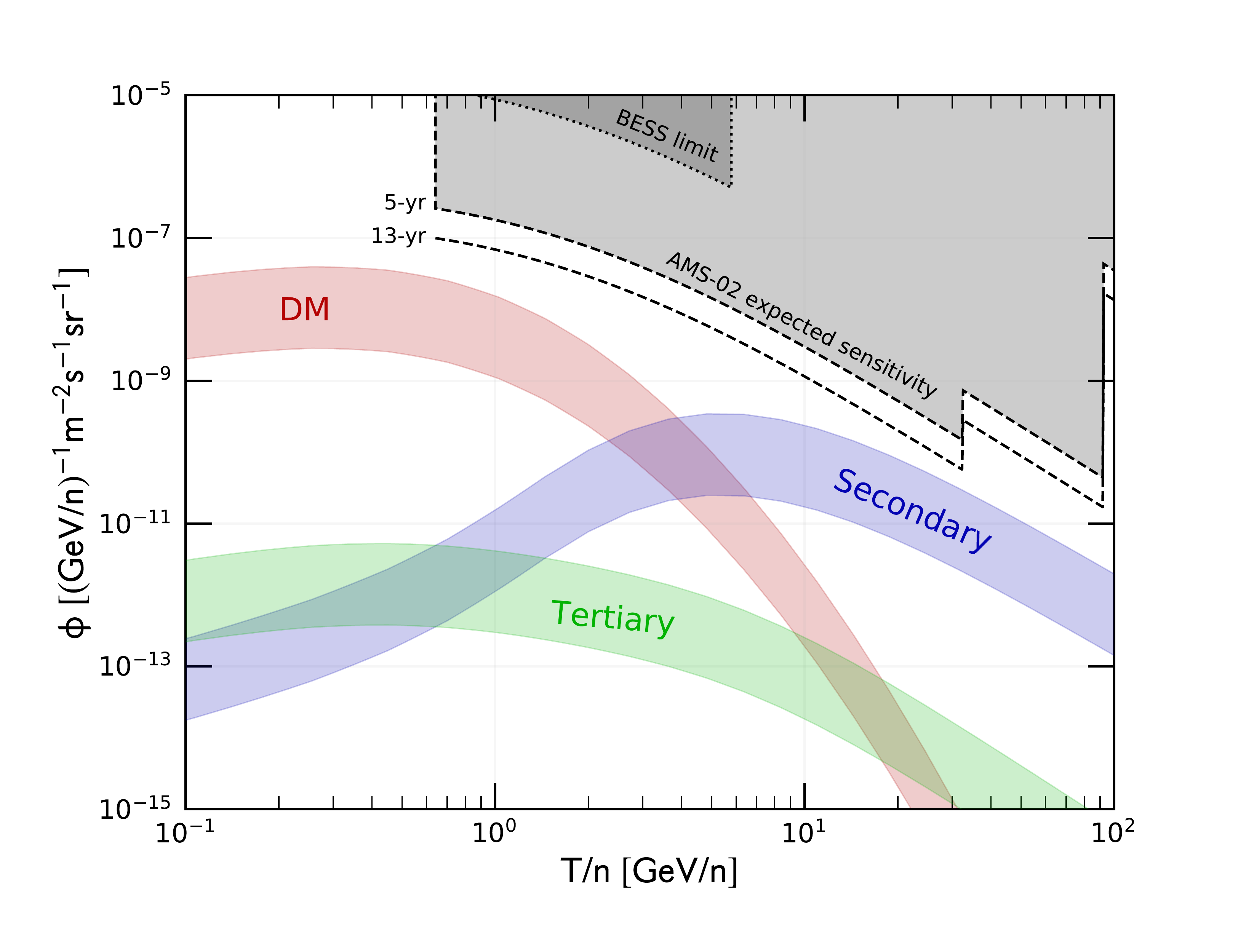}}{\includegraphics[width=0.5\textwidth]{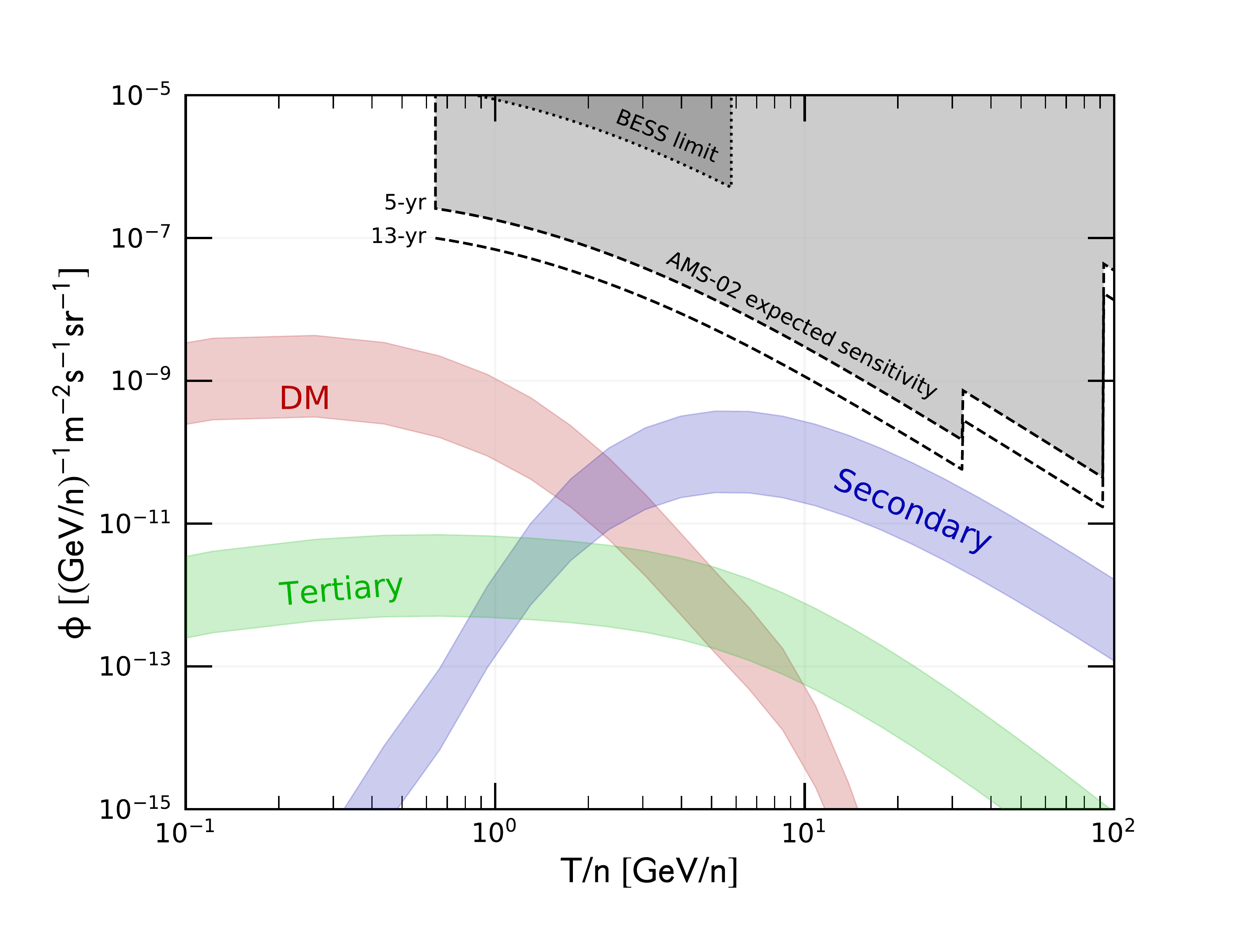}}
  \caption{  Standard astrophysical (secondary and tertiary) flux of antihelium in comparison to a potential DM signal corresponding to
             CuKrKo model. The bands show the uncertainty on the coalescence process,  
              $p_\coal $ spanning from 160 MeV to 248~MeV. The BESS limit (95\%~CL) \cite{Abe:2012tz} and 
             AMS-02 sensitivity (95\%~CL) \cite{Kounine:2011_ICRC} scaled from 18 to 5~years and 13~years
             on the antihelium-to-helium flux
             ratio are transformed to an antihelium flux sensitivity by using the measured AMS-02 helium flux.
             All lines correspond to a force-field solar modulation potential of $\phi=600$~MV, the analytic coalescence model, and
             the propagation parameters from CuKrKo (left panel) or 
             MED (right panel). 
          }
	\label{fig::antihelium_flux}
\end{figure*}

\section{\label{sec::results}Results}

\subsection{Antideuterons}
\figref{antideuteron_flux} shows the fluxes of antideuterons, separately for secondary, tertiary, and the potential DM component: the left panel refers to $\pcoal = 160$ MeV and the right panel to  $\pcoal = 248$ MeV. Furthermore, we show the different propagation scenarios with parameters taken either from the analysis CuKrKo (solid lines) or MED-MAX (shaded areas).
All shown fluxes of antideuteron are corrected for solar modulation effects by adopting the force-field approximation and assuming a Fisk potential of 400~MeV. 
Since GAPS is expected to take a balloon flight during a time period of low solar activity, this is a sound value, which is reached during most periods of solar minimum \cite{Ghelfi:2016pcv, Tomassetti:2017hbe}. In any case, the effects of different solar modulation potential on our result are mild as discussed below.
The secondary flux peaks at kinetic energies per nucleon of 3 to 4~GeV/n and quickly falls below 1 ~GeV/n and above 10~GeV/n \cite{Donato:1999gy}. The size of the flux actually does not depend on the specific propagation model above 2~GeV. At lower energies the CuKrKo propagation  leads to a larger flux compared to the analytic case MED/MAX. This is understood as the effect of reacceleration and energy losses, which are included in CuKrKo but not in our implementation of MED/MAX. The tertiary flux is suppressed compared to the secondary flux by about two orders of magnitude, but it extends towards lower energies. It exceeds the secondary flux below $0.4$~GeV/n.

The DM signal corresponds to the best fit in CuKrKo, namely annihilation into $b\bar{b}$-quark final states, a dark matter mass of 71~GeV, and a thermally averaged cross section of $2.6\cdot10^{-26}$~$\mathrm{cm^3/s}$.  As expected, its relevance is manifest in the lowest part of the energy range, below 1 GeV/n and peaks at energies between 0.1 and 0.2~GeV/n. 
For a coalescence momentum of 160~MeV the peak flux of $1\cdot10^{-5}$~$\mathrm{(GeV/n)^{-1}m^{-2}s^{-1}sr^{-1}}$ is clearly above the level of the most recent estimate of the sensitivities of GAPS (99\%~CL)~\cite{Aramaki:2015laa} and AMS-02 \cite{Aramaki:2015pii}. For the MED/MAX propagation setup, the signal is also within the detection range of both experiments.
We notice that the MAX propagation setup for the CuKrKo DM specification is probably incompatible with antiproton data \cite{Fornengo:2013xda}, but we show it for the sake of completeness.
For the larger coalescence momentum of 248~MeV, as very recently suggested by the ALICE measurements, all fluxes are up-scaled by a factor of 4 (right panel of \figref{antideuteron_flux}). Consequently, all the DM curves are well within the GAPS and AMS-02 detection range. 
\figref{antideuteron_uncertatinty} shows the reach capabilities of GAPS for the whole 2$\sigma$ allowed regions derived in CuKrKo for the DM particle compatible with the antiproton hint. The areas in the four panels of \figref{antideuteron_uncertatinty} are derived by a full scan of the DM mass and annnihilation rate inside the 2$\sigma$ 
regions and show, as a function of the DM mass, the ratio between the calculated antideuteron flux (averaged over the GAPS energy bin) and the GAPS expected sensitivity of $2.0 \times 10^{-6}$~$\mathrm{m^{-2}s^{-1}sr^{-1}(GeV/n)^{-1}}$, as determined in Ref. \cite{Aramaki:2015laa}. The GAPS sensitivity is obtained by considering two types of events in the detector (events originated by stopping antideuterons and 
in-flight annihilation events), for which the number of events required to obtain a 99\% CL detection is 1 (stopping events) and 2 
(in-flight annihilation). Whenever the ratio shown in  \figref{antideuteron_uncertatinty} is above 1 implies that GAPS will detect the corresponding antideuteron flux with a 99\% CL  confidence. This implies that the number of detected events is 1 if the detection occurs in the stopping channel, or 2 if the detection happens in the category of in-flight annihilation.
In \figref{antideuteron_uncertatinty}, the blue contour corresponds to our baseline scenario, namely the analytic coalescence model with $\pcoal=160$~GeV, solar modulation in the force-field approximation with a potential of 
$\phi=400$~MV, and propagation parameters taken from CuKrKo. We see that the whole CuKrKo parameter space would produce a detectable signal in GAPS. 
The different panels then show the changes arising from different assumptions, always compared with the baseline scenario (blue contour). Panel (a) investigates the impact of a Monte Carlo based coalescence, for which we have used the results of~\cite{Fornengo:2013osa}. This Monte Carlo approach is also tuned to ALEPH data. Note that coalescence momenta are different in the analytical and Monte Carlo approach when tuned to the same data. The signal strength drops by a factor of 4 such that the signal would be at the very edge of detectability. The larger coalescence momentum obtained from ALICE enhances the fluxes considerably and consequently the contour gets boosted: this is shown in panel (b) (again for the analytic coalescence model) where the corresponding contour for $\pcoal = 248$ MeV is pushed to a few tens of events in GAPS. This would imply several detected antideuterons. Notice that also the Monte-Carlo-based coalescence, if normalised to ALICE,  would likely imply that all of the DM parameter space is under reach of GAPS (the tuning of the Monte-Carlo-based models on ALICE requires a dedicated analysis, in order to derive its specific value for $\pcoal$, and it is not available at the moment).
Finally, the impact of solar modulation and of different CR transport models are shown in panel (c) and (d), respectively, for the analytic coalescence model. In all cases, the DM parameter space compatible with the antiproton hint is testable by GAPS. 
Notice, that the local DM density does not provide an extra uncertainty for the results of our analysis, since the annihilation rate is totally degenerate with the DM density: the DM fit in CuKrKo determines $\sv \times \rho_\odot^2$, which is the same quantity that enters in the determination of the antideuteron flux in \eqnref{sourceTerm_DM} and \eqref{eqn::NFW}.

Up to this point we considered only the case of DM annihilation into a $b\bar b$ pair. However, also other final states provide a good fit to the antiproton 
excess~\cite{Cuoco:2017rxb}. In \figref{antideuteron_final_states} we show the result for pure annihilations into two gluons ($gg$), 
$Z$-bosons ($ZZ^*$), Higgs-bosons ($hh$), or top-quarks ($t\bar{t}$). 
For the $Z$-boson we take into account that one of the two bosons might be produced off-shell\footnote{
This requires an extension of the tables in~\cite{Cirelli:2010xx} already used in~\cite{Cuoco:2017rxb}.
}, which is denoted with a star superscript. For all the channels, the DM parameter space can be tested by GAPS through antideuterons. 

Another potential indication for DM is the observed excess in gamma-rays from the Galactic center (GCE). 
Its energy spectra and morphology are compatible with a DM signal as observed and confirmed by several 
groups \cite{Goodenough:2009gk, Calore:2014xka,Calore:2014nla, Fermi-LAT:2017yoi} (and references therein). 
However, also an astrophysical explanation by unresolved point sources \cite{Calore:2014oga,Bartels:2015aea,Cholis:2015dea,Fermi-LAT:2017yoi}, 
especially millisecond pulsars, might explain the excess.  
Notice that the DM interpretation of the GCE and the cosmic antiproton excess point to very similar, compatible $m_\DM$ and $\sv$ for all 
standard model final states~\cite{Cuoco:2017rxb}. In this sense, our analysis shows that also the DM interpretation of the GCE 
is in the reach of antideuteron sensitivity for GAPS and AMS-02.

\subsection{Antihelium}
Finally we investigate the antihelium channel, for which we follow the methods introduced in Ref. \cite{Cirelli:2014qia}
and we extend the results to derive also the tertiary component. For antihelium, the coalescence momentum plays an even stronger role, since the antihelium flux is proportional to its sixth power (as compared to the third power in the case of antideuterons). Consequently, the larger coalescence momentum suggested by the recent measurement of $B_3$ in ALICE implies an antihelium flux increase by a factor of 14, as compared to the original determinations \cite{Cirelli:2014qia,Carlson:2014ssa}. The thick bands in \figref{antihelium_flux} show in fact this uncertainty on $\pcoal$. Similarly to antideuteron, we explore different propagation scenarios of CuKrKo (left panel) and MED (right panel). In the most optimistic scenario of a large coalescence momentum, the secondary antihelium flux is only a factor of 2 below the expected AMS-02 sensitivity after 13 years: this occurs for kinetic energies per nucleon of 30~GeV/n. In contrast to this, the expected signal from the DM hint is always significantly below AMS-02 sensitivity. Nevertheless, \figref{antihelium_flux} emphasizes the fact that the ability to detect low-energy antinuclei offers the best chances to identify an exotic signal, possibly originated by DM annihilation. The secondary flux is in fact  strongly suppressed below 8 GeV/n and the tertiary component does not contribute much to the background for DM particles even if the annihilation cross section was 2 order of magnitude below thermal one.

In general, the interpretation of a potential antihelium signal in AMS-02 strongly depends on the energy range of the observation. If antihelium were observed below 1~GeV/n it would be a strong indication for DM, while antihelium at energies above a few GeV/n would hint towards a determination of the secondary flux.

\begin{table}[t!]
\caption{  Summary of the best-fit DM mass and thermally averaged cross section 
          for various standard model final states from the analyses~\cite{Cuoco:2016eej, Cuoco:2017rxb}. }
\label{tab::dm_candidates}
\begin{tabular}{lcc}
 \hline \hline
 Final state    & $m_\mathrm{DM}$~[GeV]  & $\sv~[10^{-26}~\mathrm{cm^3/s}]$  \\  \hline
  $gg$          &       34             &       1.9    \\
  $b\bar{b}$    &       71             &       2.6    \\
  $ZZ^*$        &       66             &       2.4    \\
  $hh$          &      128             &       5.7    \\
  $t\bar{t}$    &      173             &       3.8    \\ \hline \hline
\end{tabular}
\end{table}

\section{\label{sec::conclusion}Conclusion}

Antimatter provides a powerful tool to indirectly investigate DM in CRs. 
We examined here the possible hint for DM annihilation in AMS-02 data on cosmic antiprotons, exploring the potential DM candidates 
with masses  from below 30 to above 200~GeV, annihilating into various standard model final states. 
We calculated the astrophysical (secondary and tertiary) as well as the DM fluxes of antideuteron and antihelium.
We found that the corresponding flux in antideuterons is within the sensitivity range of 
GAPS and AMS-02 for most of the considered scenarios. 
This conclusion has been tested against different nuclear fusion approaches and parameters, as well as propagation models and solar modulation effects. 

Along with antideuterons, we also gave predictions for the corresponding  CR antihelium, computing the primary DM flux and the secondary and tertiary components arising from interactions with the ISM.
Compared to antideuteron,  antihelium gives a similarly 
good separation of the DM signal from the astrophysical tertiary flux. However,  even in the most optimistic scenarios
the DM flux is still one order of magnitude below the AMS-02 sensitivity, while the secondary antihelium flux 
is only a factor two below the 13-year sensitivity of AMS-02. 

We stress that there is still a huge uncertainty in modeling antimatter coalescence, on the one hand, 
between applying an analytic and a Monte Carlo based model and, on the other hand, in the choice of the 
coalescence momentum. The very recent measurements of the $B_2$ and $B_3$ parameters by ALICE hint towards 
a larger coalescence probability than considered previously, increasing all the fluxes and therefore also potential signals closer to or into 
the experimentally detectable range. 
Finally, we notice that the hint of the DM signal was found at energies where the antiproton AMS-02 data are provided with an extremely high accuracy, while the interpretation is affected by 
sizeable theoretical  uncertainties. 
It is also possible that the potential DM hint simply overfits small fluctuations of the data. 
Therefore, a more {\it conservative} approach is to consider the potential signal as an upper limit on DM annihilation. Henceforth, the antideuteron and antihelium results obtained in this analysis would indicate an estimate of the highest possible fluxes without violating antiproton data.

\acknowledgements
We would like to thank Jan Heisig for providing the antiproton energy spectra for off-shell $W$ and $Z$~boson final states and Marco Cirelli for useful comments. 
This work 
is supported by the ``Departments of Excellence 2018 - 2022'' Grant awarded by
the Italian Ministry of Education, University and Research (MIUR) (L. 232/2016).
Furthermore, we acknowledge support from the research grant TAsP (Theoretical Astroparticle Physics) funded by the Istituto Nazionale di Fisica Nucleare (INFN).

\bibliographystyle{apsrev4-1.bst}
\bibliography{bibliography}{}

\end{document}